\documentclass[12pt,preprint]{aastex}

\usepackage{psfig}

\def\arcsec{$\,^{\prime\prime}$~}
\def\arcmin{$\,^\prime$~}

\def\erg/cm2sec{ergs~cm$^{-2}$~s$^{-1}$}  
\def\ergcm2{ergs~cm$^{-2}$}  
  
\def\mdot{$\dot{m}$~}

\newcommand{\lsim }{{\lower0.8ex\hbox{$\buildrel <\over\sim$}}}
\newcommand{\gsim }{{\lower0.8ex\hbox{$\buildrel >\over\sim$}}}

\def\apj{ ApJ}
\def\aap{ A\&A}

\def\pasp{PASP}

\def\aj{AJ}
\def\apjs{ApJ Supp}

\def\simge{\mathrel{%
   \rlap{\raise 0.511ex \hbox{$>$}}{\lower 0.511ex \hbox{$\sim$}}}}
\def\simle{\mathrel{
   \rlap{\raise 0.511ex \hbox{$<$}}{\lower 0.511ex \hbox{$\sim$}}}}

\newcommand{\Msun}{\ifmmode {M_{\odot}}\else${M_{\odot}}$\fi~}
\newcommand{\Rsun}{\ifmmode {R_{\odot}}\else${R_{\odot}}$\fi}

\shorttitle{LMXB and Faint Sources in NGC 6652}
\shortauthors{Heinke et al.}

\begin{document}
\title{Identification of the LMXB and Faint X-ray Sources in NGC 6652} 

\author{C. O. Heinke, P. D. Edmonds, J. E. Grindlay}
\affil{Harvard-Smithsonian Center for Astrophysics \\ 
60 Garden Street, Cambridge, MA  02138\\ cheinke@cfa.harvard.edu,  
edmonds@cfa.harvard.edu, josh@cfa.harvard.edu}

\begin{abstract}

We have detected three new x-ray point sources, in addition to the known
low-mass x-ray binary (LMXB) X1832-330, in the globular cluster NGC 6652
with a Chandra 1.6 ksec HRC-I exposure.  Star 49 ($M_{V}\sim$4.7),
suggested by Deutsch et al.(1998) as the optical candidate for the LMXB,
is identified ($<0.3$\arcsec) not with the LMXB, but with another, newly detected
source (B).  Using archival HST images, we identify ($<0.3$\arcsec) the
LMXB (A) and one of the remaining new 
sources (C) with blue variable optical counterparts at $M_{V}\sim$ 3.7 and 5.3 respectively.  The other new source (D)
remains unidentified in the crowded cluster core.  In the 0.5-2.5 keV
range, assuming a 5 keV thermal bremsstrahlung spectrum and
$N_{H}=5.5\times10^{20}$, source A has intrinsic luminosity
$L_{X}\sim 5.3\times10^{35}$ ergs s$^{-1}$. Assuming a
1 keV thermal bremsstrahlung spectrum, B has $L_{X}\sim 4.1\times10^{33}$
ergs s$^{-1}$, while C and D have $L_{X}\sim 8\times10^{32}$ ergs
s$^{-1}$.  Source B is probably a quiescent LMXB, while source C may
be either a luminous CV or quiescent LMXB.  

\end{abstract}


\maketitle

\section{Introduction}

Globular clusters have proven to be an excellent place to study
stellar populations, dynamical evolution, and binary systems.  Their
high stellar densities allow their populations of white dwarfs and neutron 
stars to interact with cluster primordial binaries by exchange
collisions (Hut et al.\ 1992).  Stellar evolution and dynamical
hardening of compact binaries within globulars lead to mass transfer
and accretion onto the compact object, which is visible especially in
the x-ray regime.  Thus x-ray studies of globular clusters are
particularly effective at studying both compact objects and compact binaries.
   
  Successful
searches for  LMXB optical counterparts have been undertaken for
several clusters (Deutsch et al.\ 1998b and references therein, Homer
et al.\ 2001).  However, of the 12 globular cluster LMXBs in the Galaxy
(cf.\ Grindlay 1993), at least 6 remain optically unidentified--those
in Terzan 1, 2, 5, and 6, Liller 1, and NGC 6652.  Of the separate
population of low-luminosity x-ray sources (Hertz \& Grindlay 1983), many 
sources have now been 
positively identified with CVs, as in NGC 6397 (Edmonds et al.\ 1999
and references therein) and with CVs, millisecond pulsars, and BY
Draconis binaries in 47 Tucanae (Grindlay et
al.\ 2001, and references therein).  

The globular cluster NGC 6652 includes one previously known LMXB,
X1832-330. This cluster is at galactocentric radius 2 kpc and 9.0$\pm$0.4 kpc
from the Sun, and is 11.7$\pm$1.6 Gyr old, using $(m-M)_V$=15.15,
$E(B-V)$=0.12 (Chaboyer 2000). The LMXB
was probably detected by {\it HEAO 1} at a low level (Hertz \& Wood
1985).  It was securely detected by ROSAT during the all-sky survey
(Predehl et al.\ 1991), again during pointed ROSAT observations
(Johnston et al.\ 1996), and in type 1 x-ray bursts by both ASCA and
BeppoSAX (in't Zand et al.\ 1998, Mukai \& Smale 2000).  A search for
the optical counterpart of the LMXB was undertaken by Deutsch,
Margon, and Anderson (1998; hereafter DMA98), unfortunately not
including the entire ROSAT error circle for the LMXB.  They were able
to identify a plausible candidate (star 49); however, it is located
2.3 $\sigma$ from the 
ROSAT position, and is identified here with a lower-luminosity source.
This paper  
presents observations with {\it Chandra} as part of a campaign to
measure precise positions for the remaining unidentified LMXBs.  We
also use {\it Hubble Space Telescope} archival images to identify
optical counterparts. 

\section{Observations and Optical Identifications}

We observed NGC 6652 on May 23, 2000 with the High
Resolution Camera--Imager (HRC-I) camera (Murray et al. 2000) on the
Chandra X-ray Observatory (CXO).  The Chandra mirrors
give well-defined point-spread functions of half-power diameter
0.76\arcsec (Jerius et al.\ 2000), oversampled by the HRC-I, allowing relative positional accuracy
to less than 0.3\arcsec$\!$.  The aspect reconstruction
systems allow absolute positional accuracy of 0.6\arcsec ($1 \sigma$;
Aldcroft et al.\ 2000). 

Data were analyzed using the {\it Chandra} X-ray Center CIAO software \footnote{
All CXC software is available at http://asc.harvard.edu/ciao/} and
standard processing, producing 1679 seconds of good data. 
The CXC source detection routine WAVDETECT (Dobrzycki et al.\ 1999) was
used to detect sources in the exposure-corrected
 central field (4\arcmin radius circle).  The PROS 
variability test vartst was applied to each source.  

Most of the HST data analyzed here (except for brief references to $B,
V$, and $U$ data) are archival data from the September 1997 WFPC2
observations of Chaboyer (GO-6517). Under this program, an orbit of short
(23s) F555W and (20s) F814W images were obtained at two different pointings,
followed by an orbit of 12 long (160s) F555W images and an orbit of 12
long (160s) F814W images. The second and third orbits were obtained at a
fixed pointing with the center of NGC 6652 placed in the center of the
PC.  
The 12 deep F555W and F814W images were cleaned of cosmic rays and combined
into single images. Aperture photometry and PSF-fitting were used to
derive instrumental m$_{555}$ and m$_{814}$ magnitudes, which were
then converted
into $V$ and $I$ using the prescription given in Holtzman et
al. (1995). The precision of the PSF-fitting photometry along the main
sequence was inferior to that obtained for aperture photometry, but was
much less affected by crowding, especially near cluster center.
HST positional information makes reference to the Guide Star Catalog v.1.1,
which has known absolute positional errors in the southern celestial
hemisphere of 1\arcsec to 1.6\arcsec
(Taff et al.\ 1990).

Six sources were detected in the central 4\arcmin radius circle using
WAVDETECT. Two sources 
more than 2\arcmin from the core are probably not associated with the
cluster, as the cluster half-mass radius is 0.65\arcmin, and the core
radius is 0.07\arcmin (June 22, 1999 version of catalog described in
Harris 1996).  We were unable to find 
counterparts for these sources, and ignore them in this paper.
We detect 4 sources within 30\arcsec of the cluster center,
including the bright LMXB 
X1832-330, hereafter source A (Figure 1); the others we label B, C,
and D. All four sources are
well-detected: A has 9140 counts, B has 75 counts, C and D have 15 and
16 respectively, with formal significances greater than 6 $\sigma$ in
each case.  We consider all four of these sources 
to be associated with the cluster, since the probability that any of 6
sources within a 4\arcmin circle would fall within 30\arcsec of the cluster
center is 9\%.  We give sources A, B, C, and D the formal designations
CXOGLB J183543.6-325926, CXOGLB J183544.5-325939, CXOGLB
J183545.7-325923, and CXOGLB 
J183545.6-325926 respectively, but use the letter designations for
this paper.  

The second brightest source (Source B) is detected in the HRC-I data
at a
position 0.8\arcsec from the position of star 49 (in the HST Guide
Star Catalog
frame). Star 49 was suggested by DMA98 as the 
counterpart for the LMXB, which we detect 
 17\arcsec away.  Star 49 has measured $(U-B)=-0.9$
(DMA98), and shows 
variability of 1 magnitude, possibly 
with a 43.6 minute periodic
modulation (Deutsch et al 2000).  DMA98 searched a 35\arcsec $\times$
35\arcsec HST WF/PC1 field (not including source A; the GO-6517 data
was not public then), for $U$-bright stars.  They   
found only one object, star 49, with $U-B$ color and magnitude similar to
known LMXBs.  Adding the GSC 1.6\arcsec (maximum)
and HRC 0.6\arcsec 
absolute positional errors in quadrature gives a 1.7\arcsec 1$\sigma$ error
circle. The 1.7\arcsec error circle is not crowded (being $\sim4$
core radii from the center), so we
confidently identify star 49 with source B. 

By matching the GO-6517 and HRC coordinates for star 49 and source B,
we performed a shift of the HRC coordinate frame by -0.04263 sec in RA,
-0.609\arcsec in Dec.  The HST images then allowed the  
identification of $V-I$ excess optical counterparts for
sources A and C within 0.01\arcsec and 0.3\arcsec of the corrected HRC
positions.  The derived x-ray and optical parameters of sources A
through D are given in Table 1.

 Figure 2
shows the $V$ vs $V-I$ CMDs resulting from PSF-fitting to both HST PC
and WF data.  The likely optical counterparts for sources A, B and C are
shown, as are some other stars (shown by 
circles, all are distant from x-ray positions)
confirmed to be blue objects by visual examination of the $V$ and $I$ images.

Figure 3 shows the combined $V$ and $I$ time series for sources A and C
(the time series for source B is reported in DMA98). A zeropoint correction
was applied to the I data so that $<V>-<I>$=0, and the errors were
calculated from the median time series rms for stars within 0.25 mag
(typically 150--400 stars) of either source A or C.  Best-fitting
sinusoids have been overplotted showing possible periodicities for A
and C.

\section{Results}

\subsection{Source A: LMXB}
 
The blue counterpart for the LMXB has $V=18.9$, $V-I=0.4$; this is 0.4
mag above the 
turnoff and 0.4 mag blueward of the main sequence in
$V-I$, similar to other known LMXBs.  Our suggested counterpart of 
source A is partly cut 
off in archival HST U images, possibly explaining why it was
not identified in DMA98.  However, the peak pixel value at the edge of the
chip clearly shows that this star is UV-bright. 

A time series analysis of the $V$ and $I$ exposures shows a clear
0.08-magnitude variability, with rms variation 4.4 $\sigma$ and
6.2 $\sigma$ in $V$ and $I$ 
respectively, where $\sigma$ is defined as the
 median rms variation of stars within 0.25 mag
of the star.  Possible periods of 0.92, or 2.22 hours
(or 4.44 hours for ellipsoidal variations), as well
as some nonperiodic flickering, are observed (see Figure 3). The
false alarm probability (FAP; cf. Scargle 1982) is $\sim0.3\times10^{-3}$ for
each period 
individually, although the short temporal coverage does not allow us
to distinguish between these periods.  We
suggest one of these is the orbital period of the system, leading 
to an orbital separation of (for M$_{total}$=1.6$\Msun$) either
$3.9\times10^{10}$ cm, $7.0\times10^{10}$ cm, 
or $1.1\times10^{11}$ cm, respectively (Frank et al.\ 1992).  Either a
low-mass M dwarf or 
a white dwarf would be a plausible secondary, depending on which
period is the real one.  

The absolute magnitude $M_V=3.7$ of the optical
counterpart is comparable to those of ultracompact ($P<1$ hr)
LMXBs in globular clusters: 3.7 for X1820-303 in NGC
6624, and 5.6 for X1850-087 in
NGC 6712 (van Paradijs \& McClintock 1994).  X-ray reprocessing into
optical light in accretion disks tends to follow the x-ray
luminosity to absolute magnitude 
scaling relation proposed by van Paradijs \& McClintock (1994, Fig. 2),
$M_V=1.57(\pm0.24)-2.27(\pm0.32)$log$\Sigma$, where $\Sigma$ is 
$(L_{X}/L_{Edd})^{1/2}(P/1 hr)^{2/3}$.  For $M_{V}$=3.7, we find
log$\Sigma=-1$ to $-0.5$ for 
periods from 0.92 to 4.4 hours.  All three proposed
periods place this LMXB within the scatter of known LMXBs.

The HRC-I has no spectral resolution, yet our flux
determination depends upon the assumed spectrum.  By using a
partial covering absorber ($2/3$ covered by $8\times10^{21}$ cm$^{-2}$),
 and a power-law with photon index 1.86, Mukai and Smale (2000) fit both
their ASCA  
GIS spectrum and the second ROSAT PSPC spectrum.  From their recorded
flux we derive 
$L_{X}$(2-8 keV)= $1.5\times10^{36}$ in the ASCA range, or
$L_{X}$(0.5-2.5 keV)= 
$6.8\times10^{35}$ ergs s$^{-1}$ extrapolated to the ROSAT range.  The fitted
partial covering absorber suggests variability in the $N_H$ column, which
would also explain the differences in spectral fitting by ROSAT
data (Johnston et al.\ 1996, Verbunt et al.\ 1995).
BeppoSAX observations (in't Zand et al., 1998) revealed Type I bursts
(observed also by Mukai \& Smale),
and persistent emission at $L_{X}$(2-8 keV)= $5.2\pm1.0\times10^{35}$
ergs s$^{-1}$ (adjusted for the change in distance to 9.0 kpc, without
absorption).  The low-level emission they observed agrees with both
ROSAT observations of low-level luminosity extended to the harder band, for
a photon index of 1.86.  

We use a power-law spectrum  with photon index 1.86, and a range of $N_H$
between $5.5\times10^{20}$ and $4\times10^{21}$, the 
first being the average cluster value and the second the maximum value
for A from ASCA observations.  These give intrinsic
luminosity ranges of $L_{X}$(0.5-2.5 keV) $\sim5.1\times10^{35}$ to 1.1$\times10^{36}$
ergs s$^{-1}$, similar to the ROSAT observations in 1990 and 1992 (Verbunt et
al. 1995, Johnston et al.\ 1996),
and $L_{X}$(2-8 keV) $\sim 5.3\times10^{35}$ to $1.2\times10^{36}$
ergs s$^{-1}$, 
comparable to the ASCA and BeppoSAX observations in 1996-7 (Mukai \&
Smale 2000, in't Zand et al.\ 1998).
We see no significant variability during the 1.6 ksec
observation.  Mukai and Smale note that
no eclipses have been seen, and that we directly observe the
neutron star during the Type I bursts, ruling out a high-inclination
geometry.  This suggests that the optical variability may be due to
orbital variations of the heated face of the secondary.

\subsection{Source B: qLMXB}

The x-ray luminosity of source B, $L_{X}$(0.5-2.5 keV) $=
4.1\pm0.5\times10^{33}$ for a 1 keV bremsstrahlung spectrum, suggests a
quiescent LMXB since CVs generally do not reach this
luminosity (Verbunt et al.\ 1997). Source B is the only source to show
clear x-ray variability, at the 95\% level in the Cramer-Von Mises
test (Daniel 1990).  It is possible that this source  
is responsible for the Type I bursts seen with ASCA and BeppoSAX (in't
Zand et al.\ 1998, Mukai \& Smale 2000), since
neither telescope could resolve source B from the bright LMXB, source A,
only 17\arcsec distant.

DMA98 identified star 49 as the bright LMXB counterpart based on U-B excess
($U-B=-0.9$, 1.2 mags blueward of the main 
sequence [MS]), and
later (in Deutsch et al.\ 2000) added weight to the identification by
discovering $\sim1$ magnitude flickering and a possible
period of 43.6 minutes. We confirm the variability, at 24 $\sigma (V)$
and 21 $\sigma (I)$, and identify the
same possible optical period, but with an FAP
of 0.027, the least significant FAP of the power spectrum peaks for
the three optical counterparts.

If the 43.6-minute period is the real binary period,
a main sequence companion is ruled out since the implied orbital
separation for a 
 2 $\Msun$ system is $3.6\times10^{10}$ cm, less than the radius of even an M2
 star, implying a double-degenerate system.  However, this
identification suffers a problem in that star 49 
 falls onto the MS in $V-I$ (see Figure 2, Table 1). The system also
lies on the MS in an instrumental $V$ vs. $B-V$ CMD from
further archival analysis.  The MS colors suggest that light from a
main-sequence secondary is dominant in $V$ and $I$, and that the weak
accretion disk provides the $U$ excess (see DMA98). However, this
interpretation cannot explain why both $V$ and $I$ vary in tandem by
one magnitude (Deutsch et al.\ 2000, Figure 1) without invoking
eclipses, nor does it explain the large flickering (which generally
implies a bright, blue disk).  
 
We considered the possibility that the identification suffers
source confusion (background or foreground stars contributing $V$, $I$
light), but there is no evidence for blending in the HST
image.  We consider it most likely that the 43.6 minute period is
spurious, but the 1-mag variability remains a mystery.

\subsection{Source C: CV or qLMXB?}

The suggested counterpart of source C lies only 3"
from the cluster center, $\alpha$=18:35:45.66,
$\delta$=-32:59:26.5 (J2000), from Grindlay et al.\ 2001 (in
preparation). This is within one core radius (4.2\arcsec, Harris 1996).  Its 
$V-I$ color, $M_{V}$ , and $F_{X}/F_{V}$ (see Table 1, Figure 2)
suggest a CV, since light from field CVs is usually dominated by the hot
accretion disk throughout the optical 
range.  Although both the x-ray luminosity and $M_{V}$ are
 bright for a CV, 
bright magnetic CVs (IPs; see Barrett et al.\ 1999), or CVs in
outburst, reach this range.  Archival GO-6095 HST WFPC2 data
shows C at similar brightness and color, and it is unlikely that
we have caught C in outburst twice.
A quiescent LMXB is the other 
plausible interpretation.  However, the accretion disk in a
low-\mdot system is expected to be truncated due to a mass transfer
instability leading to an ADAF flow, and the light from the secondary
should be dominant for $V-I$. 

The optical time series for source C show clear variability (see
Figure 3), at the 5.0 $\sigma(V)$ and 7.6 $\sigma(I)$ levels.  We find
a possible periodicity of 44.4 minutes  
(close to the 48-minute HST half-period),
semi-amplitude 0.16 magnitudes, and 
false alarm probability of $0.88\times10^{-2}$ ($\sim1$\%). Three other
stars within a 3$\sigma$ error circle (indicated by crosses in Figure 2) show no variability at the $2\sigma$ level.  The time series for C
shows significant flickering, which
might combine with the periodic HST window to generate the possible
44.4 minute periodicity.  This period is disturbingly close to the suggested
43.6 minute period for source B, indicating that both periods may be spurious.
    
\subsection{Source D: Multiple sources?}

The final source, D, lies less than 0.4\arcsec from
the cluster core (as given above).  The x-ray flux shows no
significant variability.  A bright blue star ($V$=17.4, $V-I$=0.2) is
located 0.7\arcsec away from the corrected x-ray position (see Figure
1), but shows no variability (significance$\lsim2\sigma$).
Based on its nonvariability and magnitude (see Figure 2), we judge it
to be a blue straggler, and unlikely to be the counterpart.  Accurate
photometry of source D's 
error circle down to the expected magnitude of CVs or qLMXBs ($m_{V}\gsim20$)
may be impossible due to the overlapping psfs from 
nearby blue stragglers and red giants.  

It is quite possible that source D is really multiple sources
overlapping within the central core, where the concentration of close
binaries and thus x-ray
sources is expected to be high. Possibilities for source D include
a qLMXB, a bright millisecond pulsar, and/or one or more CVs.

\section{Discussion}

Our detection limit in this observation was $L_X$(0.5-2.5 keV)$\sim4\times10^{32}$ ergs s$^{-1}$,
for a 90\% confidence detection threshold 4-count source, background
 being 0.04 counts arcsec$^{-2}$ at the core.   Most 
detected field CVs are at $\lsim10^{32}$ ergs s$^{-1}$ in this band, and would thus be missed by this short
exposure.   The total
 detected emission from within one core radius, assuming a 1 keV
thermal bremsstrahlung spectrum, was $1.5\times10^{33}$ ergs s$^{-1}$, and
can be accounted for by the two detected sources.  A deeper ACIS
exposure could measure spectra, confirm the possible 43.6 minute
period for source B, and 
identify low-luminosity sources such as CVs.  Several potential CVs or
hot white dwarfs on the WF2 and WF4 chips are indicated in
Figure 2.   Additional blue non-variable stars are noted on the PC chip; their
lack of variability and blue colors resemble the non-flickerers noted in
NGC 6397 and proposed to be He white dwarfs (Edmonds et al.\ 1999).
However, these could be field stars.  Using the figures from
Ratnatunga and Bahcall (1985) for Galactic star counts in the
direction of NGC 6652, we estimate that we should
see $\sim$1.2 stars with $B-V<0.8$ between $19<m_V<23$ in the PC
field, and $\sim$11 stars with $B-V<0.8$ between $21<m_V<25$ in the
combined WF2 and WF4 fields.  Thus the WF2 and WF4 blue stars could be
explained by field contamination, but it is unlikely that all the blue stars 
in the PC are field stars. 
 
The locations of the sources in this cluster are intriguing.  Sources
C and D are within one core radius of the center, but A and B (each
expected to be 
$\sim$1.6-2\Msun) are at least 6 and 4 core radii respectively from the
center of the 
cluster.  They are not in dynamical equilibrium in the cluster, since
binaries tend to sink toward the cluster core within the local
relaxation time (Hut et al.\ 1992).  The half-mass relaxation time for
NGC 6652 is $3.5\times10^{8}$ years, and both sources are
well within the half-mass radius of 39\arcsec (Harris 1996). 
We consider the possibility that they have been ejected from the
cluster core in a separate paper which considers the radial
distribution of all 12 bright LMXBs using both our new Chandra
positions and updated cluster parameters (centers and core
radii; Grindlay et al.\ 2001, in preparation).

\acknowledgments

This work was supported in part by Chandra grant GO0-1034A.
The ability to make this observation is a testament to the dedication
and hard work of the {\it Chandra} team.


\vspace*{-4cm}
\hspace{-3cm}

\begin{deluxetable}{cccccccc}
\tablewidth{6.9truein}
\tablecaption{X-Ray Source Parameters}
\tablehead{
\colhead{Src} 
 & \colhead{RA$^{a}$} & \colhead{Dec} &
\colhead{$F_{X}$\thinspace$^{b}$} &  \colhead{$F_{X}/F_{Opt}$\thinspace$^{c}$} &
\colhead{$L_{X}$\thinspace$^{d}$}
& \colhead{$M_V$}
 & \colhead{$V-I$}  
}
\startdata

A  &  18:35:43.649$\pm$.001 & -32:59:26.77$\pm$.01 &
$5.9\!\times\!10^{-11}$  &   16 & $1.1\!\times\!10^{36}$ & 3.7 & 0.34      \\
B  &  18:35:44.543$\pm$.009 & -32:59:38.97$\pm$.09 &
$3.7\!\times\!10^{-13}$ &   0.3  & $4.1\!\times\!10^{33}$ & 4.7 & 0.85      \\
C  &  18:35:45.751$\pm$.011 & -32:59:23.51$\pm$.06 &
$7.3\!\times\!10^{-14}$ &   0.1 & $8.4\!\times\!10^{32}$  & 5.3 & -0.06      \\
D  &  18:35:45.643$\pm$.012 & -32:59:26.74$\pm$.15 &
$7.3\!\times\!10^{-14}$ &  -    & $8.4\!\times\!10^{32}$ &  -     & -   \\

\enddata
\tablecomments{
$^a$ The errors are statistical centroiding errors; 
systematic errors on the absolute coordinates are 0.6\arcsec
(1$\sigma$), from Aldcroft et al. 2000.  Coordinates are shifted onto
the HST Guide Star Catalog frame as described in text.
$^b$ For 0.5-2.5 keV band, received flux in ergs s$^{-1}$ cm$^{-2}$,
assuming 1 keV thermal bremss 
spectrum for sources B, C, and D, power law of $\alpha$=1.86 for A, 
$N_{H}=5.5\times10^{20}$ for B, C, and D and $N_{H}=4\times10^{21}$
for A. \quad  
$^c$ Ratio of x-ray (0.5-2.5keV band) to
optical/UV flux (using log $F_{Opt}=-0.4V-3.96$, from Verbunt et al.\
1997).\quad
$^d$ Intrinsic luminosity in ergs s$^{-1}$, 0.5-2.5 keV, assuming
$N_H$ and spectra as above, d=9.0 kpc, 
$(m-M)_{V}=15.15$, from Chaboyer et al.(2000).  
}
\end{deluxetable}


\clearpage

\hspace{-2cm}
\vspace*{4cm}
\psfig{file=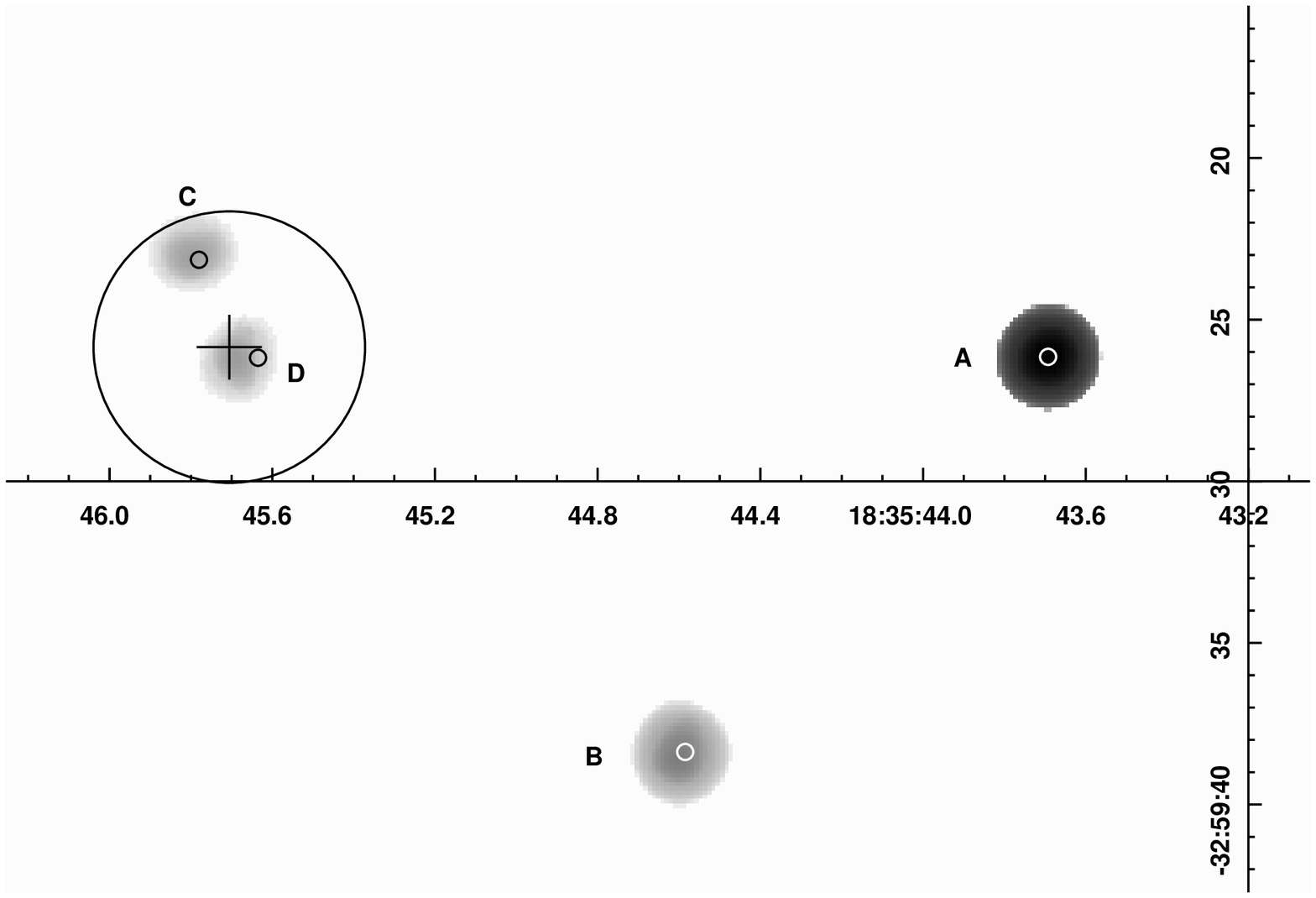}

\vspace*{-4cm}

\figcaption[figure1.ps]{HRC-I Chandra image of NGC 6652.  
Image smoothed by wavelet detection
algorithm WAVDETECT, part of Chandra analysis software package CIAO.  
The large circle represents one core radius (4.2\arcsec, Harris 1996),
and the cross marks the location and estimated uncertainty in the
cluster center (Grindlay et al.\ 2001, in preparation).  The small
circles (radius 0.25\arcsec, WF/PC2 FWHM for point sources) represent
the locations (shifted to HRC frame) of the three 
likely optical counterparts, as well as the blue star near source
D. \label{Figure 1}}

\hspace{-3cm}
\vspace{-3cm}
\psfig{file=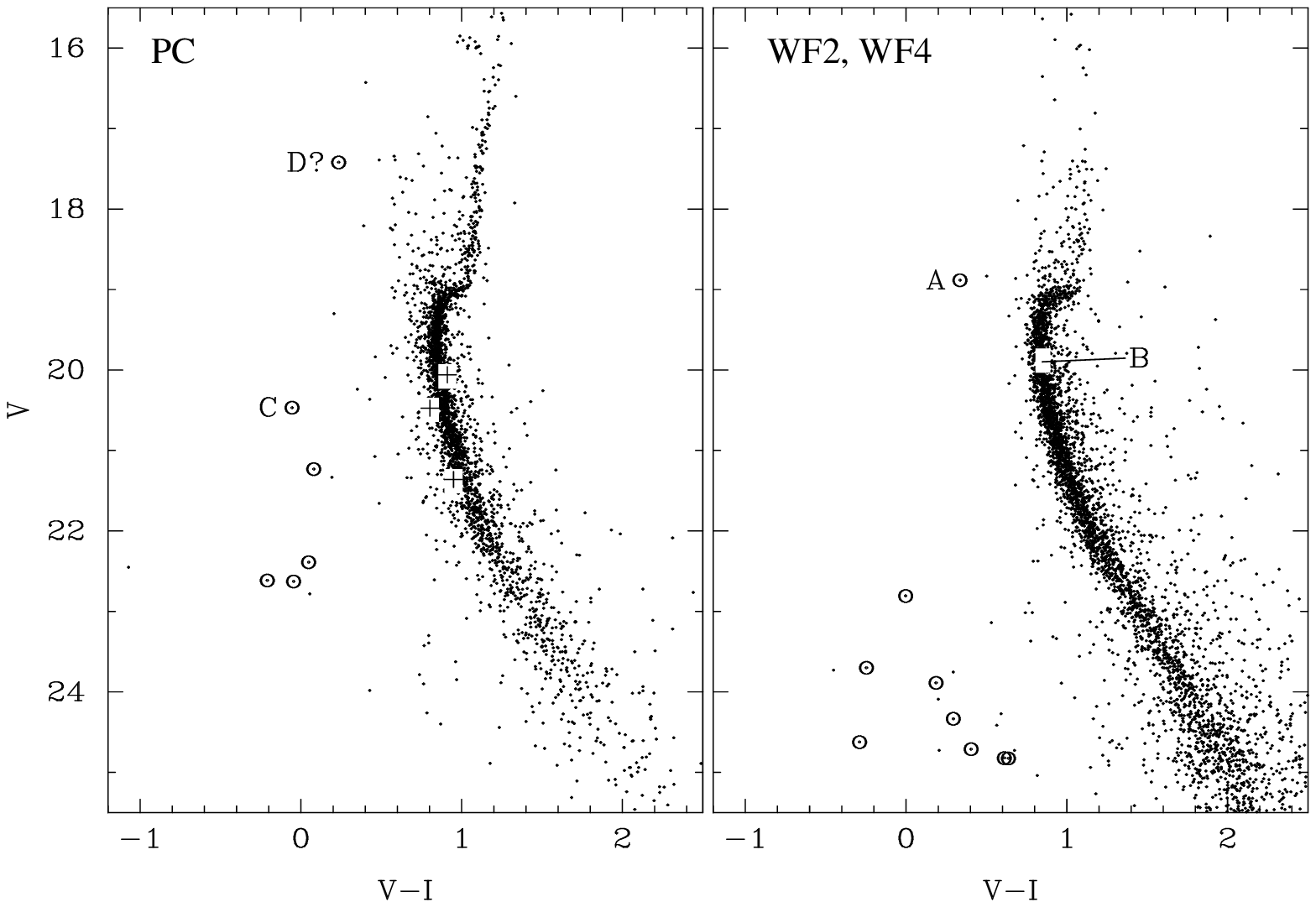}

\vspace{-4.5cm}
\figcaption[figure2.ps]{V-I color-magnitude diagrams of NGC 6652, 
showing three probable counterparts (A, B, C) of
detected x-ray sources.  The left panel
shows the CMD for the PC and the right panel shows the CMD for WF2 and WF4
combined (note that undersampling and saturation affect the giant branch
in the WF CMD).  Note that source B falls on the main sequence
in V-I.  A bright blue nonvariable star near the location of source D is
indicated as D?. Crosses mark the location of three normal stars also
falling within the $3\sigma$ error circle for C.  Several blue stars are
also indicated by circles, see text. \label{Figure 2}} 

\hspace{-3cm}
\vspace*{0.1cm}
\psfig{file=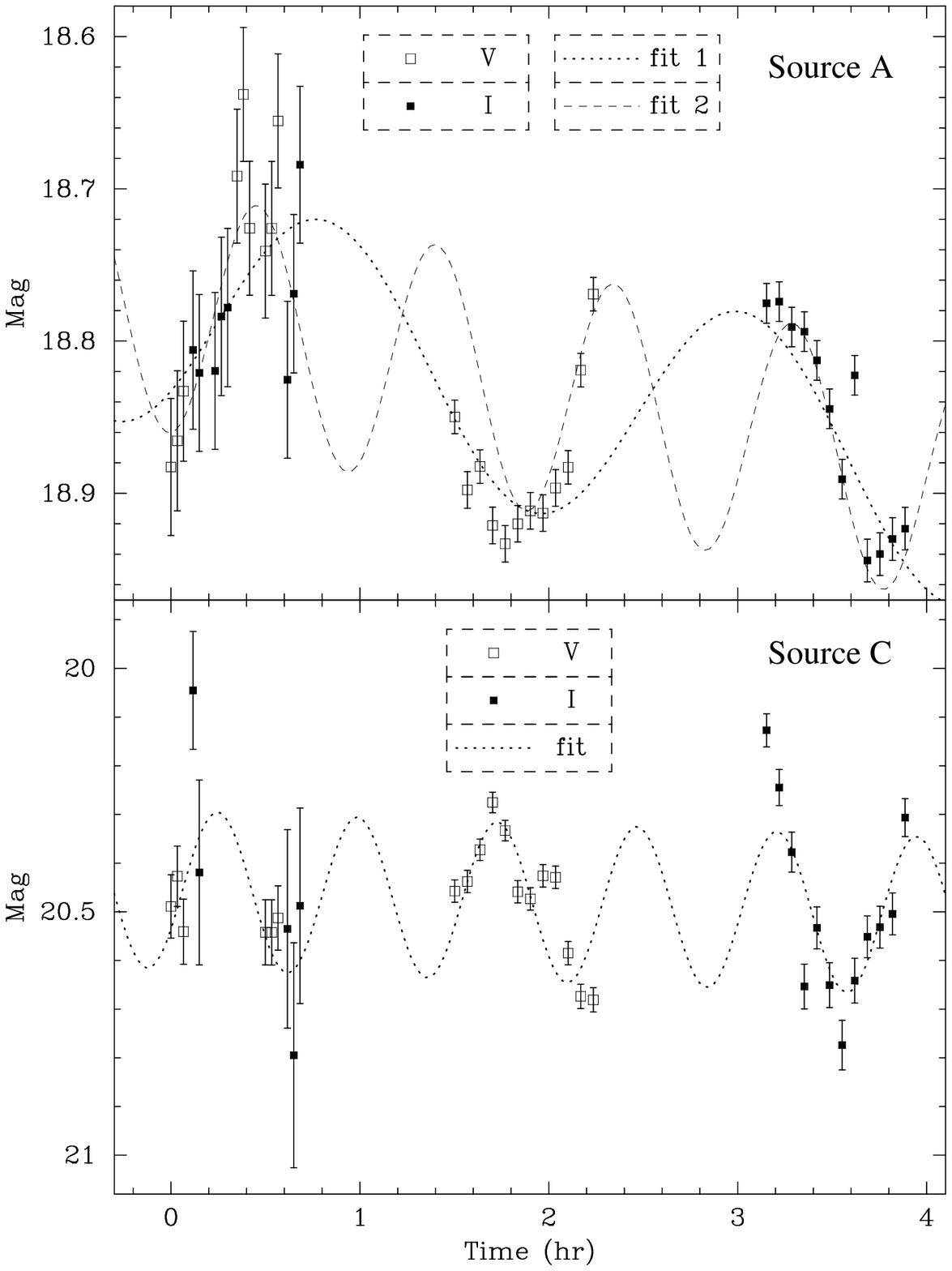}
\vspace*{-4cm}

\figcaption[figure3.ps]{Optical light curves for sources A and C
from three HST orbits.  The open squares represent $V$ data and the
filled squares
$I$ data, shifted to correct for time-averaged color.  Two 
sinusoidal fits are plotted for source A, with periods of 0.92 (fit 2) and
2.22 (fit 1) hrs.\label{Figure 3}}

\end{document}